\documentclass[journal,comsoc,10pt]{IEEEtran}
\usepackage[export]{adjustbox}
\usepackage{textcomp}
\usepackage{algorithm}
\usepackage{algpseudocode}
\usepackage{tabularx}
\usepackage[table]{xcolor}
\usepackage{booktabs}
\usepackage{array}
\newcolumntype{P}[1]{>{\centering\arraybackslash}p{#1}}
\newcolumntype{M}[1]{>{\centering\arraybackslash}m{#1}}
\usepackage{multicol}
\usepackage{multirow}
\usepackage{caption}
\usepackage{subcaption}
\usepackage{color}
\usepackage{capt-of}
\usepackage{balance}
\usepackage{dsfont}
\usepackage{graphicx}
\usepackage{epstopdf}
\usepackage{stfloats}
\usepackage[cmintegrals]{newtxmath}
\usepackage{amsfonts}
\usepackage{times}
\usepackage{latexsym}

\usepackage{amssymb}
\usepackage{amsmath}
\usepackage{verbatim}
\usepackage{bm}
\usepackage{cancel}
\usepackage{wrapfig}
\usepackage{listings} 
\usepackage{url}
\usepackage{xcolor}
\usepackage{lettrine}
\usepackage{cite}
\usepackage{pifont}
\usepackage{diagbox}

\renewcommand{\IEEEQED}{\IEEEQEDclosed}
\graphicspath{{figure/}{figures/}}
\IEEEoverridecommandlockouts

\begin{document}
\title{Multi-armed Bandits for Link Configuration in Millimeter-wave Networks}
\author{Yi Zhang, \IEEEmembership{Member,~IEEE} and Robert W. Heath Jr., \IEEEmembership{Fellow,~IEEE}
	\thanks{
		This work was partially supported by the U.S. Army Research Office under grant W911NF1910221, NSF grants CNS-1731658 and CNS-1910112. Yi Zhang is with The University of Texas at Austin, TX 78731, US (e-mail: yi.zhang.cn@utexas.edu). Robert W. Heath Jr. is with North Carolina State University, Raleigh, NC 27695, US (e-mail: rwheathjr@ncsu.edu).
	}}
\markboth{}
{}
\maketitle

\begin{abstract}
Establishing and maintaining millimeter-wave (mmWave) links  is challenging due to the changing environment and the high sensibility of mmWave signal to user mobility and channel conditions. MmWave link configuration problems often involve a search for optimal system parameter under environmental uncertainties, from a finite set of alternatives that are supported by the system hardware and protocol. 
For example, beam sweeping aims at identifying the optimal beam(s) for data transmission from a discrete codebook. Selecting parameters such as the beam sweeping period and the beamwidth are crucial to achieving high overall system throughput. 
In this article, we motivate the use of the multi-armed bandit (MAB) framework to intelligently search out the optimal configuration when establishing the mmWave links.
MAB is a reinforcement learning framework that guides a decision-maker to sequentially select one action from a set of actions. As an example, we show that within the MAB framework, the optimal beam sweeping period, beamwidth, and beam directions could be dynamically learned with sample-computational-efficient bandit algorithms.
We conclude by highlighting some future research directions on enhancing mmWave link configuration design with MAB.
\end{abstract}


\section{Introduction}\label{sec:Intro}
Achieving the highest performance in a wireless communication network requires constant readjustment to changing channel and network conditions. Such optimizations happen across all layers over multiple timescales to achieve end-to-end performance objectives. Many of these algorithms involve successive operations of adaptation and reconfiguration. For example, beamforming in a millimeter wave network might involve frequent beam training (finding the best beam in a beam-training codebook~\cite{Javid_active_learning}) and a less-frequent reconfiguration (finding the best beam codebook size among several possible codebooks~\cite{TSCB_YiZhang_Mobihoc_2021}). Solving the reconfiguration problem is challenging as the optimal solution changes over time due to dynamics in the environment and mobility in the network, and it is unrealistic to find the optimal solution at each configuration point due to large training overhead. Sequential decision-making, a type of reinforcement learning, is one approach for formulating and solving such problems. 

In this paper paper, we motivate the use of the multi-armed bandit (MAB) framework~\cite{MAB_Auer_JML_2002} to solve sequential decision-making problems in mmWave and subTHz wireless communication networks. Such problems occur frequently when operating a mmWave link in a higher mobility setting as found in mobile ad hoc networks, vehicular networks, and cellular networks. 
A typical MAB framework consists of a decision-maker (aka agent) that sequentially selects one action (aka arm) from a set of actions. A scalar reward would be revealed to the agent given the selected action. 
The typical challenge in a MAB problem is to design an appropriate arm selection algorithm with optimality and convergence.
MAB has been successful in many applications of reinforcement learning such as assessing treatment effectiveness in clinical trials~\cite{bandit_app_clinical} and predicting user preference in recommender systems~\cite{bandit_app_recommendation}. 

More recently, MAB has been applied to solve problems in wireless communications. Relevant examples include
learning and tracking a rate and a MIMO mode pair that provides highest data rate in 802.11 systems~\cite{bandit_app_rate_adaptation}, finding an optimal set of beams for beam training in mmWave networks~\cite{FML_BA_contextual_bandit_V2I_Infocom_2018,BA_OSUB_Hashemi_Infocom_2018}, and selecting the optimal beamwidth for mmWave beam training in our prior work~\cite{TSCB_YiZhang_Mobihoc_2021}.
Prior work has showed how MAB can improve cumulative system throughput compared to reasonable non-sequential solutions. 

In this paper, we show the power of MAB to solve link configuration problems in mmWave communication networks related to configuring the beamforming weights of the transmit and receive antenna arrays. We explain how selecting the beam sweeping period, the beamwidth of the beams in the codebook and even the candidate beams in the codebook can be formulated as sequential decision-making problem. We introduce key background on MAB to explain how they can be used to solve these problems. We provide some examples of performance in terms of system throughput in a mobile mmWave network. Finally, we highlight some potential future research directions for exploiting MAB in mmWave system designs. While we focus the description and simulation study on mmWave communication at frequencies below $100$ GHz, the ideas can be applied to subTHz by adjusting the system parameters like antenna array size and updating the channel models. 

\section{Sequential decision-making in mmWave link configuration}
Beamforming using antenna arrays is critically important for mmWave communication~\cite{Overview_mmWave}. The provided array gain obtained with well configured antennas overcomes losses due to shrinking antenna size and wider communication bandwidths. In IEEE 802.11ad/ay and 5G, the defacto approach for configuring those antenna arrays is through beam sweeping, more generally known as beam training. Though there are many variations, the key idea is for the transmitter and receiver estimate the link performance with different combinations of transmit and receive beam pairs, and then to exchange information and agree on the best pair for communication.

Generally speaking, there is a tension between the codebook size and the net system throughput. In the absence of side information, achieving a high array gain typically requires time to search over candidate beams from a larger codebook, which takes away from time used for communication with the optimum pair. If the beam coherence time is long, as in a fixed wireless scenario, that cost may be sufficiently amortized. But if the beam coherence time is short, as in a vehicular ad hoc network, it may take too long to find the optimum beam configuration leaving little or no time for communication. The problems are exacerbated by uncertainty in the system due to network dynamics, user mobility, channel conditions and blockages. For example, beam training may have to be performed more frequently for a codebook of narrow beams due to misalignment errors, potentially canceling out the benefits compared to using wider beams with less array gain. 

The codebook size and throughput dilemma can be solved in part by dynamically adjusting the system parameters to local conditions as they change. This leads to what is known as a sequential decision-making problem where the system (acting as the agent), chooses the system parameters from a set of possible parameters (the action), to achieve good average throughput (the reward). As explained in this paper, by formulating link configuration problems using the sequential decision-making framework, they can be solved efficiently using MABs. Essentially, the choice of the system configurations is adjusted over time based on estimates of performance and feedback to the agent. A good algorithm balances exploration (trying different configurations) and exploitation (using a configuration that works well). MAB is the perfect tool to solve those configuration selection problems in mmWave systems due to its sampling and  computational efficiency.

In the rest of this section, we describe three important mmWave link configuration problems including what parameters need to be optimized. Later in the paper, we provide numerical examples that give insights into the gains that can be achieved when using MABs to solve these problems. 

\subsection{Beam sweeping period optimization}
The beam sweeping period is one system parameter that may be adjusted in a typical mmWave communication system. In 5G New Radio (NR) release 15, there are five possible beam sweeping periods: $10$ms, $20$ms, $40$ms, $80$ms, and $160$ms. The beam sweeping period is the time between attempts at beam training, and the optimum value depends not just on the geometry of the environment but also the mobility of devices and objects acting as potential blockers in that environment. For example, consider a vehicle-to-infrastructure communication link on a multi-lane highway. Traveling faster on the road generally requires a shorter beam training period than going slowly down the road. That period may be further reduced if there are many trucks on the highway blocking the signal, while it could be larger if the road is empty.  Using the shortest period of $10$ms all the time would lead to the most resilient connection, but also to high overhead during the occasions that the beams do not need to be swept so frequently. In this case, the sequential decision-making problem is for the system to adapt the optimum sweeping period over time. 

\subsection{Beamwidth optimization}
The beamwidth is a parameter that may be optimized as a means to realize beam codebooks with different sizes. In line-of-sight channels, the highest array gain is achieved using beams that are narrow and matched to the array response for the direction of the angle-of-arrival or departure. It can be advantageous though to build codebooks with beams that have wider beams.  Such codebooks can be smaller since fewer beams are required to cover a given angular area. As such, they can be searched more quickly. The selected beam is also optimum over a larger angular range, perhaps increasing the beam sweeping period. The main drawback of wider beams is that they sacrifice array gain, reducing the rate achieve achieved during communication and possibly requiring a longer dwell time per beam.    

The choice of the optimum beamwidth is both scenario and device dependent. For example, wide beams benefit devices that are moving quickly (more resilient to beam pointing error) and those that are closer to the base station (do not need the array gain derived from searching a larger codebook). Any inappropriate choice could potentially degrade the overall system throughput~\cite{TSCB_YiZhang_Mobihoc_2021}. In this situation, the sequential decision-making problem is for the system to adapt the beamwidth and therefore the codebook size over time. 

\subsection{Codebook downsampling}
The propagation channel at mmWave bands has fewer strong paths compared to lower frequencies, since there are more losses due to scattering. Moreover, the devices are normally non-uniformly distributed on a given site. Taken together, these imply that among the available candidate beams, it is plausible that only a few of them are potentially feasible, while others may point to the spatial directions where mmWave propagation channels rarely exist. Identifying a small set of useful beams and using it for link configuration could greatly reduce reconfiguration overhead without significantly degrading the link quality. This allows obtaining the benefits of smaller codebook sizes along with those of higher gain narrow beams. in this case, the sequential decision-making problem is to find the codebook, selected from among the possible beams in a larger codebook.

\section{Background on MAB}
A classical MAB framework describes an agent that performs sequential decision-making in a discrete-time setting. 
At each time step, the agent must select one action from a set of actions, and then a scalar reward would be revealed to the agent given this selected arm. The provider of this reward is called the environment, which captures the overall system uncertainty. The objective of a MAB problem is to identify and use the optimal arm (the one provides the largest expected reward) as much as possible. A mmWave system could be viewed as an agent. Then, for example, different beam sweeping periods or beamwidth could be regarded as different actions in the MAB framework. Or, a set of beams or a pair of beam sweeping period and beamwidth could be also treated as an action. The reward would be the effective data transmission rate when using the selected configuration setup.

We further explain the MAB framework mathematically as follows. Supposing there are $K$ arms, the reward of the $k$-th arm is modeled as an unknown distribution $\pi_k$ with mean $\mu_k$. Note that $\pi_k$ can be stationary (time-invariant) or non-stationary and in this article, we focus on stationary $\pi_k$ for simplicity and the other case would be discussed. 
The agent may, however, make a non-optimal decision in an arbitrary time step due to the lack of information of those reward distributions, i.e. $\left\{\pi_k\right\}_{k=1}^{K}$. 
The MAB framework defines a term called regret to quantify the loss of reward when selecting a non-optimal arm. To be specific, denoting $I[t]$ as the index of the arm pulled, $reward[t]$ as the scalar reward whose expectation is $\mu_{I[t]}$, and $regret[t]$ as the associated regret generated in the $t$-th time step, then the regret $regret[t]$ is defined as the difference between the highest expected achievable reward, denoted by $\mu_{k^*}$, and the expected reward of the selected arm $I[t]$. By considering a finite-time horizon $T$, the cumulative regret, denoted by $R[t]$ is given as
\begin{equation}
R[T]  = \sum_{t=1}^{T} regret[t] = \sum_{t=1}^{T} \left(\mu_{k^*} - \mu_{I[t]}\right).
\end{equation}
The objective of the MAB framework is to minimize the accumulated regret $R[t]$ over the finite-time horizon 

The advantages of applying MAB for mmWave link configuration can be further summarized as below:
\begin{enumerate}
	\item Most tunable mmWave link configuration parameters such as beamwidth and beam sweeping period have only a finite multiple of alternatives to ease standardization, commercialization, and implementation. This perfectly matches the framework of MAB.
	\item MAB has a small sample and computational complexity, and does not rely on neural network training (NN) as deep reinforcement learning does. This makes MAB a perfect tool for wireless communication where the samples (i.e. feedback) are usually not available offline for the NN training.
	\item Operating at high-frequency bands also raises up many new hardware challenges like beam squint, phase noise and etc. The MAB framework can incorporate all these uncertainties into the unknown distribution of the reward. Hence it is more generalizable and deployable for different scenarios.
\end{enumerate}

A MAB algorithm refers to a policy that is used by the agent to decide which arm to play at the current moment given all historical actions and corresponding feedbacks $\left\{(reward[t],I[t])\right\}$. Most MAB algorithms for selecting arms are based on Upper confidence bound (UCB)~\cite{MAB_Auer_JML_2002} or Thompson sampling (TS) technique~\cite{Empirical_TS_Chapelle_NIPS2011}. Their key ideas are briefly reviewed as below: 

\subsubsection{Upper confidence bound (UCB)}
The UCB-type algorithm directly focuses on estimating the mean reward of each arm. By gradually gathering more knowledge of the arms, it adaptively moves from exploring different arms to concentrating on exploiting a certain arm.
The UCB-type algorithm measures each arm such that the true mean reward is below the upper confidence bound with high probability. At each time step, the arm with the highest upper confidence bound would be chosen. 
Typically, this upper confidence bound is designed as the sum of the empirical reward estimate and a term that is inversely proportional to the number of times the arm has been exploited. 
For example, the UCB1 algorithm~\cite{MAB_Auer_JML_2002} uses the form $\mu_i+\sqrt{\frac{2\ln(n)}{n_i}}$ where $\mu_i$ is the empirical reward estimate of arm $i$ at the current round, $n$ is the number of trials that have been performed and $n_i$ is the number of trials that were performed with arm $i$ until the current round.
Accordingly, as time evolves, a better reward estimate is achieved and the decision-making process would be primarily relying on the empirical estimate. 
There are many variants of UCB-type algorithms which revise the second term $\sqrt{\frac{2\ln(n)}{n_i}}$ for different purposes. 

\subsubsection{Thompson sampling (TS)}
TS is a Bayesian-type approach that estimates the distribution of the reward with respect to a given prior distribution. We briefly describe it as follows.
Unlike UCB-type algorithms, TS aims at selecting the arm (at each time slot) that has the largest probability of being optimal with the available historical observations. 
To be specific, TS first models the reward by a parameterized likelihood function $P(r|\theta, k, x)$ which depends on some unknown parameters $\theta$, the selected arm $k$, and some optional context information $x$ ($x$ is available before each decision making). It then assumes a prior distribution of the unknown parameter $\theta$, which is denoted as $P(\theta)$. 
At each round of decision-making, TS samples a $\tilde{\theta}$ from the distribution $P(\theta|\mathcal{D})$. It then selects the arm that has the largest expected reward given this sampled parameters $\tilde{\theta}$, namely that $\arg\max_{k=1,2,\hdots,K}\mathbf{E}\left\{r|\tilde{\theta},k,x\right\}$.
After pulling the selected arm, an observation is provided by the environment, which is denoted as a triple $\mathcal{D} = (x;k;r)$. TS then use the observation to update the prior distribution $P(\theta)$ by the Bayesian rule, i.e. $P(\theta |{\mathcal {D}}) \propto P({\mathcal {D}}|\theta)P(\theta)$, where $P({\mathcal {D}}|\theta)$ is the likelihood function which is derivable from $P(r|\theta, k, x)$.
As time evolves, the accumulated observations form an accurate approximation of the distribution of the reward, hence a better decision would be made. 
TS achieves comparative performance as the state-of-the-arts UCB policies and can provide significantly better performance for some cases~\cite{Empirical_TS_Chapelle_NIPS2011}. Another advantage of TS is that it does not need parameter tuning. It is favorable to use TS if there is prior distribution of the arms' rewards and their posterior update is computationally efficient. For example, a Dirichlet distribution could be used as a prior to model the arm that follows a categorical distribution~\cite{MultinomialTS_Riou_ICALT_2020}. 

The expected cumulative regret of both the UCB algorithm and TS-based algorithm can be upper-bounded by $O(\log T)$ under certain conditions. Please refer to~\cite{MAB_Auer_JML_2002} and~\cite{MultinomialTS_Riou_ICALT_2020} for more details. More generalization of the above MAB framework such as incorporating contextual information and extending to multi-agent MAB are found in~\cite{BanditFor5G_2016_ComMag}.

\section{Applications of MAB in mmWave link configuration}\label{app_example}
In this section, we will showcase how to use the MAB framework to make decisions among the available link configuration settings in mmWave systems. In particular, we provide link-level simulations to demonstrate the performance improvement, in terms of data rate, achieved by using the MAB framework.

\begin{figure}[!t]
	\centering
	\includegraphics[width = 0.38\textwidth]{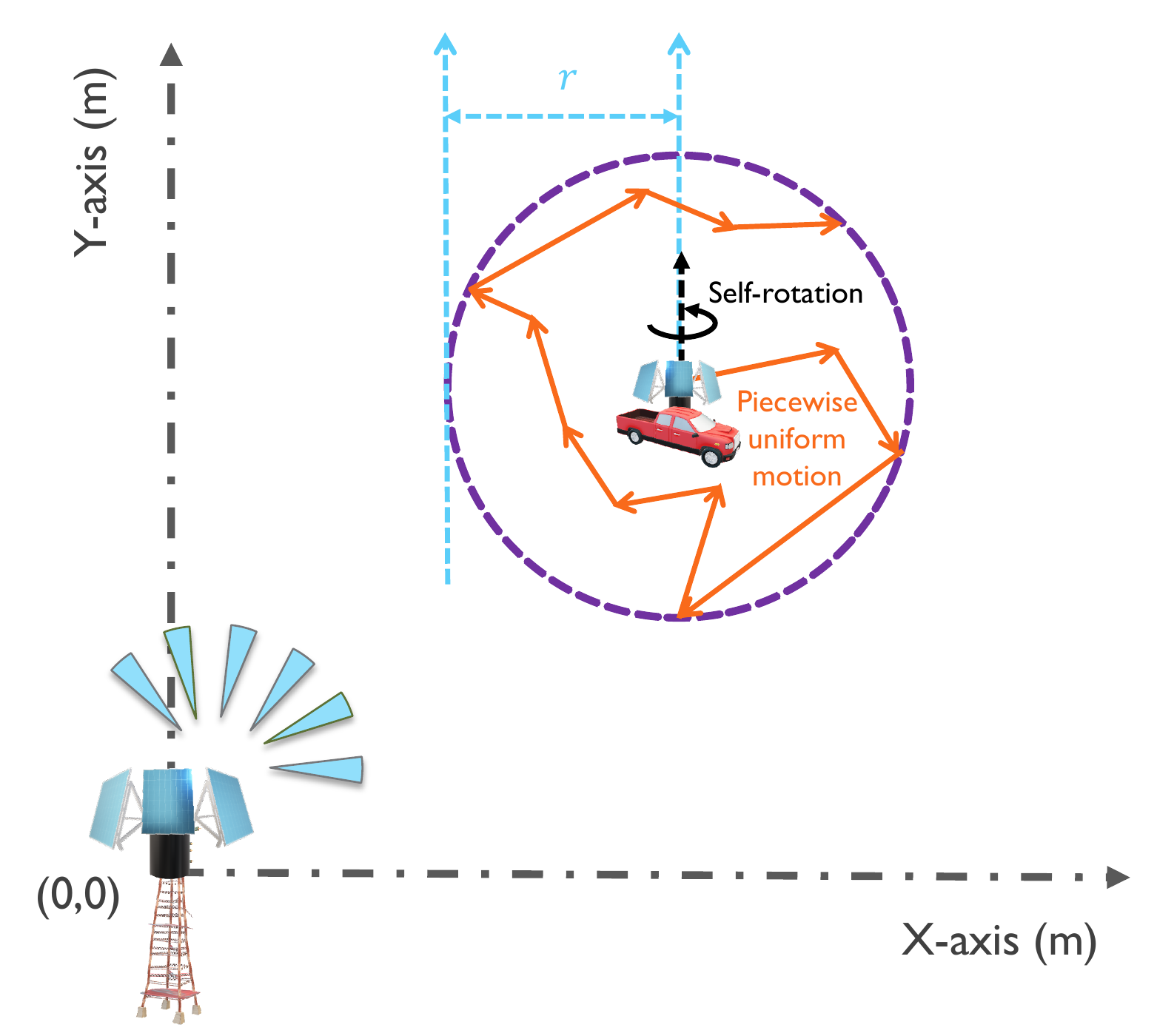}
	\caption{User randomly moves and self-rotates within a predefined area which is a circle with that the center coordinate is (21.21,21.21) m and the radius $r$ is 10 m. The initial location of the user is randomly generated within the circle.}
	\label{fig:network_topo}
\end{figure}

Our basic setup is a mmWave network where a base state (BS) (or an access point (AP)) keeps communication with a mobile user through the beam sweeping-based mechanism (2D beam sweeping is considered here). The user is moving (with random speed and direction) and self-rotating within a circular area as shown in~Fig.\ref{fig:network_topo}. 
To mimic the dynamic propagation environment, we model that the user experiences random blockage with a certain probability, meaning that its direct link to the BS randomly suffers from a signal attenuation of $PL^\text{block}$ dB. 
	
The studied mmWave system is modeled in a discrete-time setting where each communication time slot refers to a complete round. To be specific, at beginning of each time slot, the BS selects one of the available configuration setups to first perform beam sweeping and then uses the beam with the largest signal-to-noise ratio (SNR) for data transmission to the user. At the end of a time slot, the effective data rate would be calculated as the total data transmitted divided by the time slot duration. We will use MAB algorithms to select the configuration setup at each time slot.

The key simulation parameters are summarized in Table~\ref{tab:sys_paras}; some unlisted parameters would be given in the specific scenarios that are to be discussed. For more details, the source code is accessible in GitHub\footnote{Link: https://github.com/yzhang417/MAB-MmWave-Link-Config}. All the evaluation results presented below are averaged over 500 realizations.
In the following, three link configuration scenarios are investigated. 

\begin{table}[t!]
	\captionsetup{type=table,justification=centering}
	\captionof{table}[t]{Link-level simulation parameters} 
	\label{tab:sys_paras}
	\begin{tabular}{||M{4.2cm}|M{3.3cm}||}
		\hline
		\rowcolor{gray!20}
		System and channel parameter & Value \\
		\hline
		Carrier frequency ($f_c$) & 60 GHz \\
		\hline
		Bandwidth & 2.160 GHz \\
		\hline
		Duration of a beam pair measurement & 10 $\mu$s\\
		\hline
		Shadowing fading ($\chi$)  & $\chi\sim \mathcal{N}(0,2)$ dB\\
		\hline 
		Path loss model ($PL^\text{LOS}$) & $28 + 22 \log_{10}(\text{user distance}) + 20\log_{10}f_c + \chi$\\
		\hline
		\rowcolor{gray!20}
		Hardware parameter & Value \\
		\hline
		Number of arrays of AP & 4 \\		
		\hline
		Number of arrays of UE & 4 \\
		\hline
		Elevation beamwidth  & $75^\circ$ \\
		\hline
		Transmitting power of AP (dBm) & 15 \\
		\hline
		\rowcolor{gray!20}
		Blockage and mobility parameter & Value \\
		\hline
		Blockage attenuation $PL^\text{block}$ (dB) & 20 \\
		\hline
		Probability of being under blockage & 0.13 \\
		\hline
		Speed range & [5,10] m/s \\
		\hline
		Rotation rate range & [0,10] degree/s \\
		\hline
	\end{tabular}\centering
\end{table}

\subsection{Scenario I: beam sweeping period selection}
\begin{figure*}[!t]
	\begin{subfigure}[t]{0.64\textwidth}
		\centering
		\includegraphics[width = 0.99\textwidth]{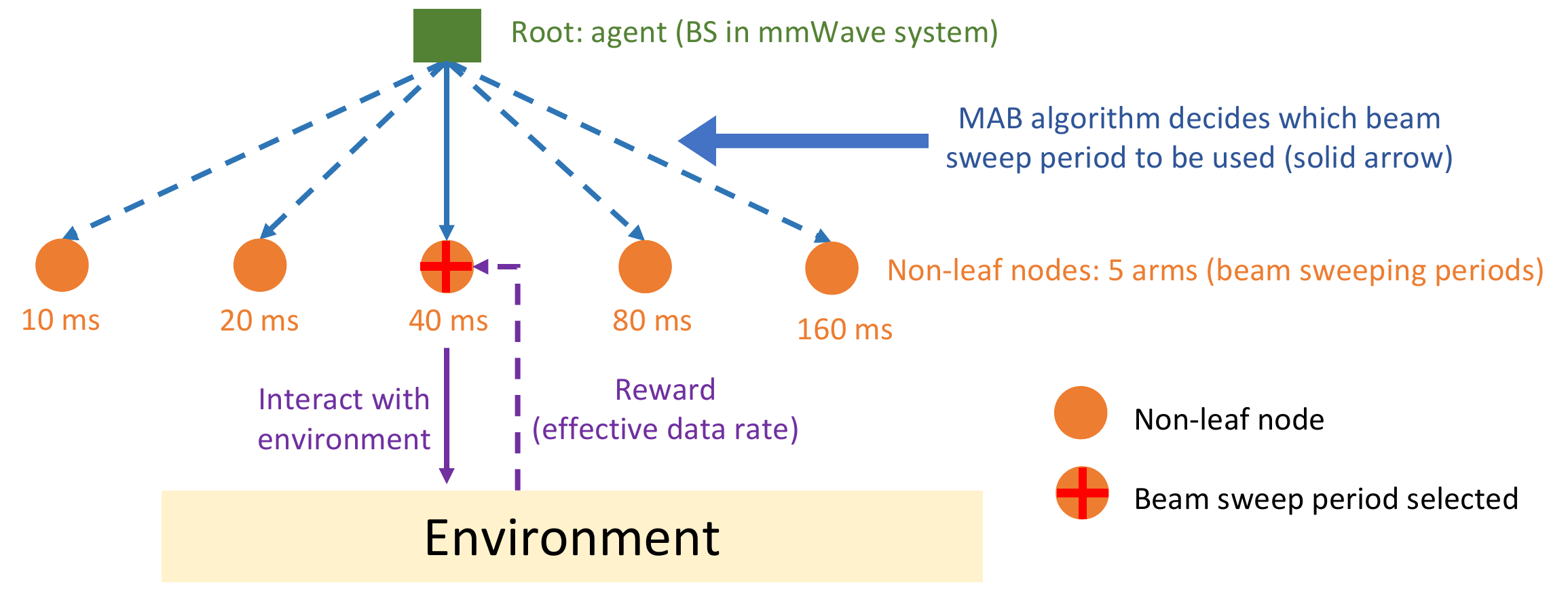}
		\caption{MAB framework adaption in scenario I (Beam sweeping period selection)}
		\label{fig:scenario1}
	\end{subfigure}
	\begin{subfigure}[t]{0.36\textwidth}
		\centering
		\includegraphics[width = 0.99\textwidth]{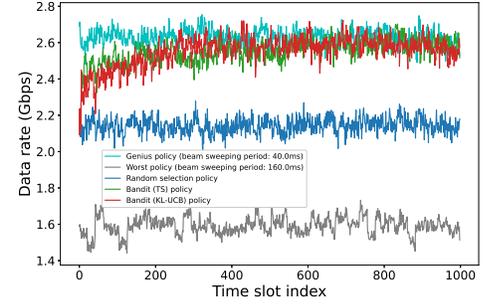}
		\caption{Evaluation result in scenario I}
		\label{fig:tslot_selection}
	\end{subfigure}
	\caption{(a) Each beam sweeping period as an arm. (b) Evolution of effective data rate versus time slot index: Each time slot corresponds to one beam sweeping period. Within 500 time slots, the bandit policies converge to the genius policy and improve the data rate by 22\% with respect to a random selection policy and by 36\% with respect to the potentially worst policy.}
\end{figure*}

For this first scenario, we consider optimizing the beam sweep period for the given basic setup. 
We suppose that five alternatives of beam sweeping periods (i.e. time slot duration) are supported by the system protocol and they are 10, 20, 40, 80, 160 ms. In the beam sweeping process, 256 sectors (around 1.41$^{\circ}$ in term of beamwidth) are used by the BS, and 16 sectors (around 22.5$^{\circ}$ in term of beamwidth) are used by the user. To apply the MAB framework, we view each beam sweeping period as an arm and the reward is the effective data rate achieved at the end of the time slot, which shown as~Fig.~\ref{fig:scenario1}.

The evaluation results of the bandit-based policies along with some benchmarks are given in Fig.~\ref{fig:tslot_selection}. Instead of showing the cumulative regret, we demonstrate the evolution of the effective data rate during the learning process since the performance improvement in date rate is the ultimate goal from the perspective of wireless context. By evaluating all the static policies that used a fixed arm throughout, the genius policy is the one that provides the highest average data rate and the worst policy is the one that results in the lowest data rate. Both TS-type and UCB-type bandit policies are evaluated.
Some interesting observations can be drawn from Fig.~\ref{fig:tslot_selection}: 
(1) The two bandit policies start by behaving like a random selection policy and converge quickly into the optimal solution. The convergence takes around 500 time slots, which corresponds to 5-80 seconds (as the slot duration could vary from 10 to 160 ms). 
(2) The bandit policies boost the system data rate by 22\% in comparison to the random selection policy. Moreover, choosing a suboptimal beam sweeping period could potentially lead to 36\% performance degradation, from around 2.5 Gbps to 1.6 Gbps (the worst policy). This well justifies the necessity of using a learning-based framework to configure the mmWave link in dynamic environments. 
(3) We can see that the genius policy uses the period 40 ms, which is primarily driven by the mobility setting given in Table~\ref{tab:sys_paras} where the user moves with an average speed of 7.5 m/s. It is safe to infer that if the user is static, the genius policy would switch to using the longest period to exploit the beam sweeping result as long as possible.
 
\subsection{Scenario II: joint beam sweeping period and beamwidth selection}
We now further consider a scenario where the used beamwidth is not necessarily fixed for the BS like that in Scenario I.
Instead, we assume that six 6 beam sets different beamwidth are available at the BS and they are sorted from wide to narrow levels, which corresponds to 16, 32, 64, 128, 256, 512 sectors, respectively. Each beam set consists of beams sharing the same width and together covering cover the entire spatial area. It is worth pointing out that the generation of beams of different resolutions could be achieved by beamforming optimization or antenna on-off techniques~\cite{TSCB_YiZhang_Mobihoc_2021}. The optimal beamwidth is determined by both user channel conditions and beam sweeping period. This motivates us to jointly configure beam sweeping period and beamwidth.

\begin{figure*}
	\begin{subfigure}[t]{0.64\textwidth}
		\centering
		\includegraphics[width = 0.99\textwidth]{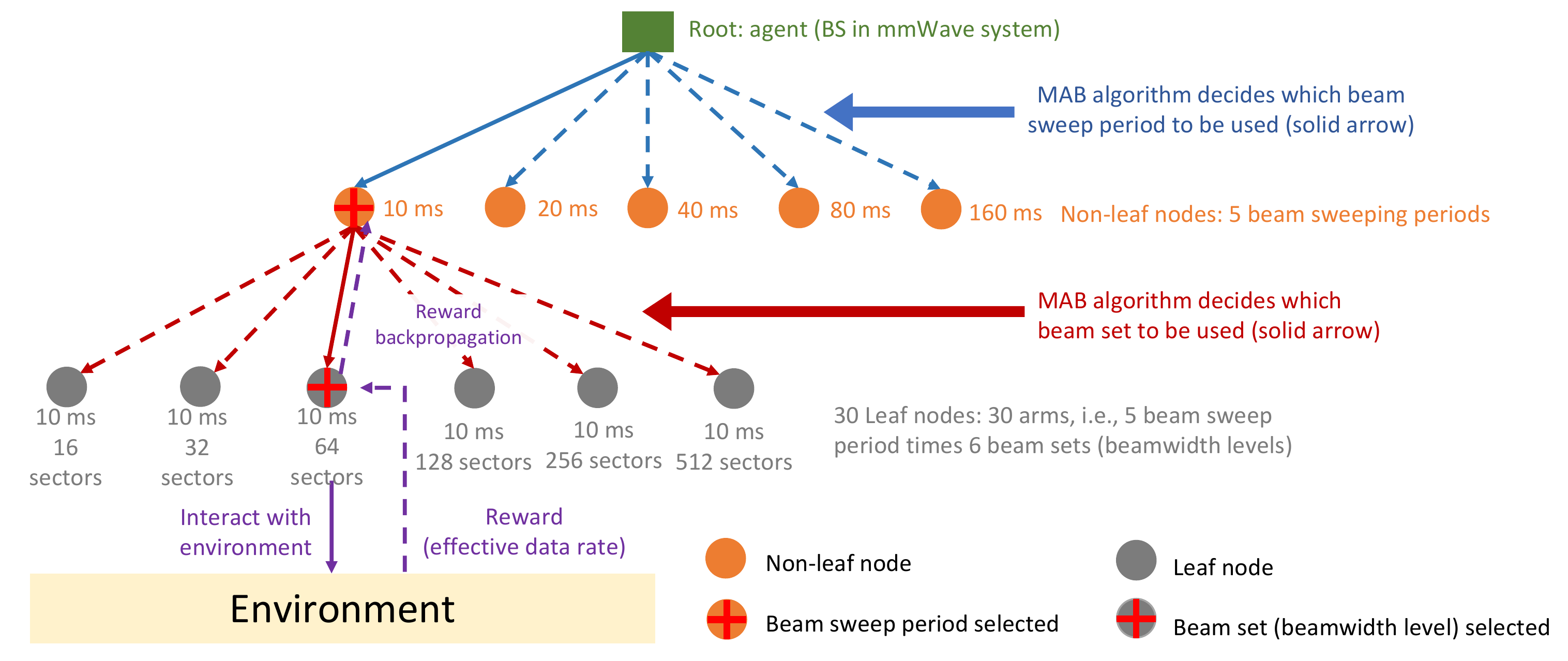}
		\caption{MAB framework adaption in scenario II (Joint beam sweeping period and beamwidth selection)}
		\label{fig:MTCS}
	\end{subfigure}
	\begin{subfigure}[t]{0.36\textwidth}
		\centering
		\includegraphics[width = 0.99\textwidth]{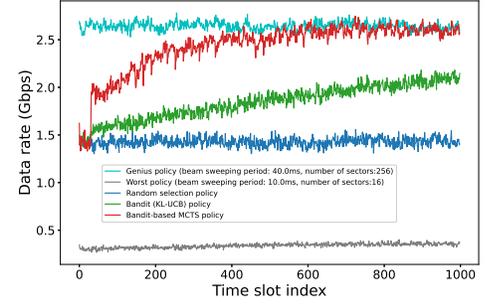}
		\caption{Evaluation result in scenario II}
		\label{fig:tslot_bw_selection}
	\end{subfigure}
	\caption{(a) Mechanism of bandit-based MCTS: Each non-leaf node represents a set of actions and it stores the empirical reward of this set of actions. All the 5 non-leaf nodes have their own 6 child nodes, though only the child nodes of the first node (10 ms) are shown as an example. Each leaf node represents a single action. One node is selected at each depth by a MAB algorithm until a leaf node is reached. The interaction with the environment only occurs at the leaf node and the reward would be backpropagated to its ancestors to update the reward estimation. 
	(b) Evolution of effective data rate versus time slot index: Each time slot corresponds to one beam sweeping period. For the bandit-based MCTS policy, the KL-UCB algorithm~\cite{KLUCB_Garivier_COLT_2011} is used to selected the nodes. Within 500 time slots, the MCTS policy converges to the genius policy and improves the data rate by 30\% with respect to a classical bandit policy, by 68\% with respect to a random selection policy and by over 1000\% with respect to the potentially worst policy.}
\end{figure*}

For this joint optimization, a straightforward application of MAB is to consider that there are 5$\times$6 arms (5 beam sweeping periods and 6 beamwidth levels) and then apply the classical MAB algorithms. This approach, however, does not exploit the fact that the action space (30 arms in this scenario) can be factored into several subsets, namely that each beam sweep period could be combined with one alternative beamwidth level to become an arm.
To exploit this special structure, another type of bandit approach decomposes the action space into multiple sequential sub-actions and sequentially decides on sub-actions, which is referred to as Monte-Carlo tree search (MCTS)~\cite{Bandit_MontCarlo}. We further explain MCTS as follows. As shown in Fig.~\ref{fig:MTCS}, each non-leaf node represents a set of actions (arms) and it stores the empirical reward of this set of actions while each leaf node represents a single action. At the beginning of each decision-making, the MCTS algorithm starts from the root node and then traverses down until a leaf node is selected. Each node on the path is selected by a MAB algorithm. The action that is associated with the selected leaf node would be executed in the environment to get a reward. Finally, the reward would be backpropagated to its ancestors to update the estimates of their arms. The intuition behind this is that MCTS could keep tracking the estimated reward of the non-leaf nodes that are encountered in the earlier sub-action selection. Therefore, if certain non-leaf nodes are re-encountered, their estimates could be used to bias the decision on the next sub-action, which may speed up the convergence.

The evaluation results of the MAB-based joint beam sweeping period and beamwidth optimization are shown in Fig.~\ref{fig:tslot_bw_selection}. We first observe that the bandit-based MCTS policy is superior to the random selection policy by providing more than 68\% data rate improvement within 500 time slots (5-80 seconds). Besides, the potential worst policy could result in a low data rate of 0.25 Gbps, which is less than 10\% of the data rate achieved by the bandit-based MCTS policy. 
This big performance gap is due to the fact that the more system configuration parameters are wrongly chosen, the worse performance it could be.
The bandit-based MCTS policy converges much faster than the classical MAB algorithm, which empirically justifies that structure of the action space should be exploited to speed up the learning process.

It is worth noting that in the above scenario, the beamwidth optimization is performed given that the exhausted beam sweeping mechanism is adopted. If a hierarchical beam-searching mechanism was used, MAB is still applicable and the corresponding beamwidth optimization would be to optimize the highest beam resolution that the hierarchical search should narrow down to.

\subsection{Scenario III: joint beam sweeping period, beamwidth, and beam direction selection}
In the above two scenarios, the beam sweeping process has probed the whole 2D angular space to find the best link. In the real system, it might not be necessary to scan the whole space when the user(s) only move around a certain location in the coverage area. For example, scanning the first quadrant in Fig.~\ref{fig:network_topo} would be sufficient to find a line-of-sight path, which could reduce the overhead by 75\%. In this last showcased scenario, which is built upon Scenario II, we would give more possibility to the beam sweep process by only testing partial directional beams in each time slot. This leads us to joint perform beam sweeping period, beamwidth, and beam direction selection.

\begin{figure*}
	\begin{subfigure}[t]{0.64\textwidth}
		\centering
		\includegraphics[width = 0.99\textwidth]{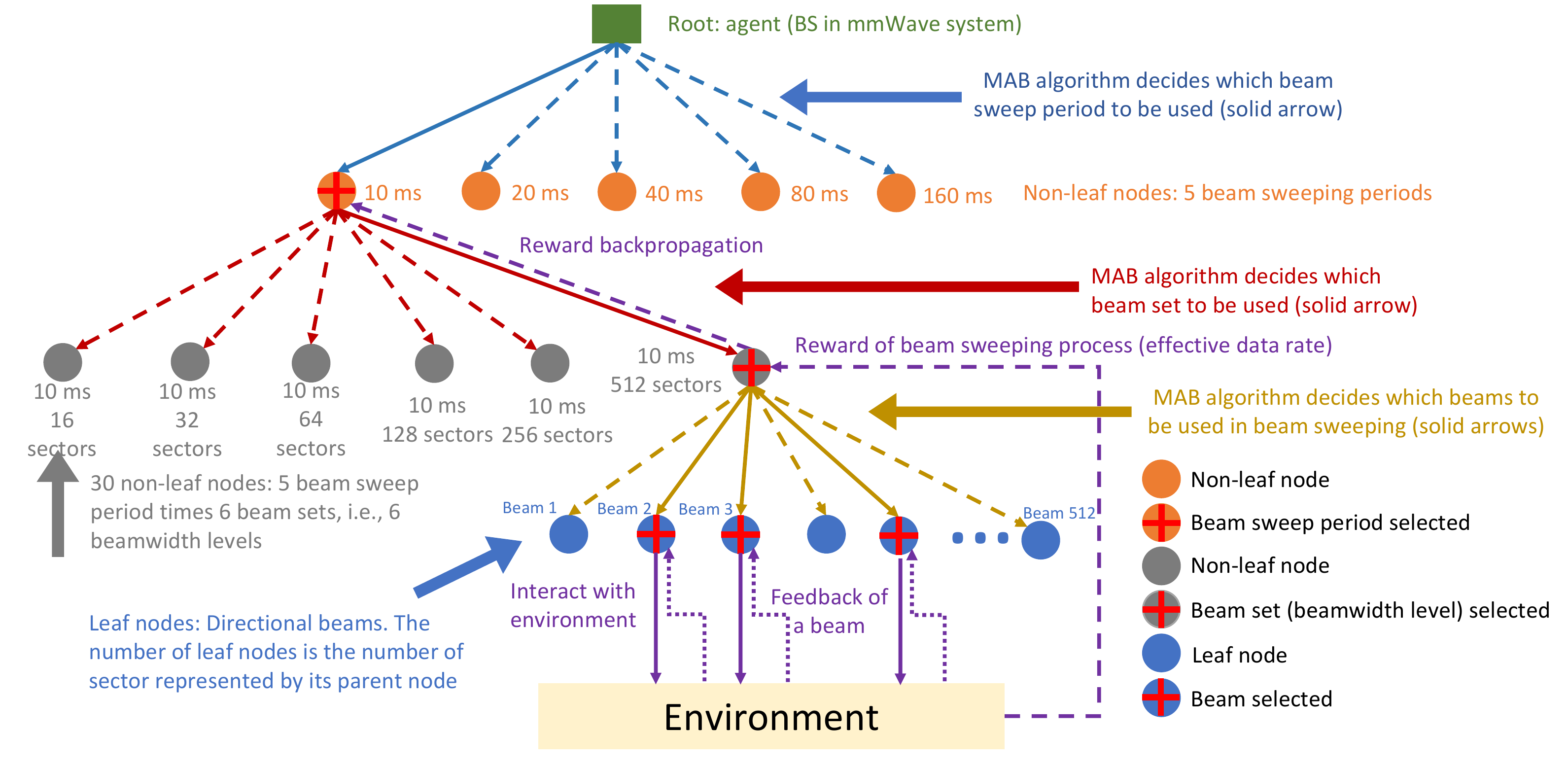}
		\caption{MAB framework adaption in scenario III (Joint beam sweeping period, beamwidth, and beam direction selection)}
		\label{fig:topKbeam}
	\end{subfigure}
	 \begin{subfigure}[t]{0.36\textwidth}
		\centering
		\includegraphics[width = 0.99\textwidth]{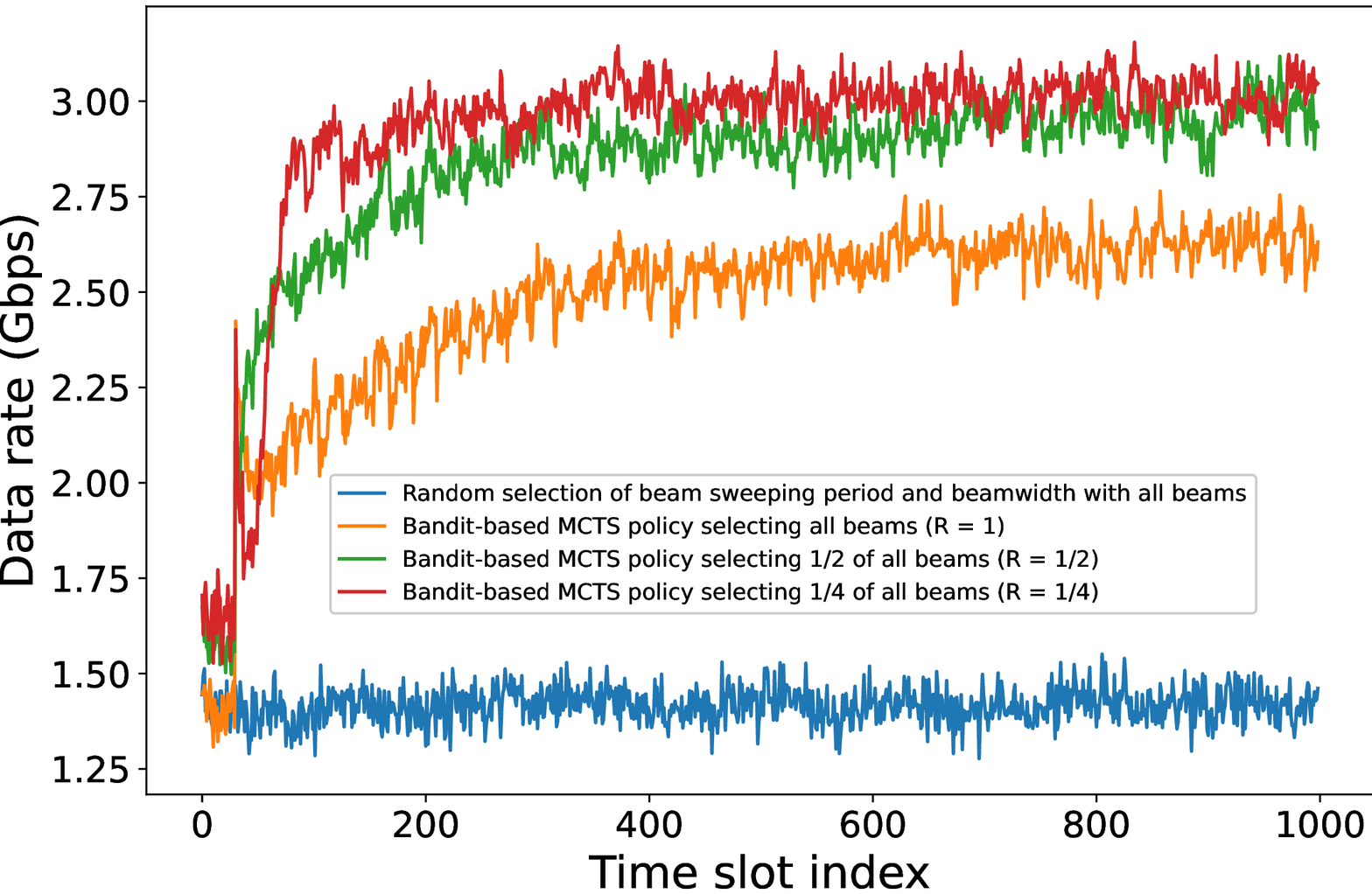}
		\caption{Evaluation result in scenario III}
		\label{fig:tslot_bw_beam_selection}
	\end{subfigure}
	\caption{(a) Adoption of best-$K$ identification into MCTS: the non-leaf node selection remains the same as that in Fig.~\ref{fig:MTCS} while multiple leaf nodes are selected to interact with the environment. Here, each gray non-leaf node has their own child nodes while the child nodes of only one node are shown as an example. Each leaf node updates its estimate with its own feedback. A reward, which is a function of the feedback of the selected leaf nodes, is backpropagated to the ancestor nodes. In our scenario, the feedback of a beam is whether it connects the user and the reward is the effective data rate after the beam sweeping process. (b) Evolution of effective data rate versus time slot index: Each time slot corresponds to one beam sweeping period. For MCTS-based policy, the beam sweeping period and beamwidth are selected using KL-UCB index while the beam directions are selected using UCB1 index. Within 300 time slots, the MCTS policy that tests half of the beams ($R=1/2$) improves the data rate by more than 20\% with respect to the MCTS policy that enumerates all candidate beams, and by more than 150\% with respect to the random selection policy.}
\end{figure*}

Given a selected beamwidth, we consider that only a port of total sectors is selected for the beam sweeping, and this portion is denoted as $R$. For example, if the beamwidth associated with 512 sectors are selected and $R=1/4$, then only 128 out of the 512 sectors would be used at each time slot for beam sweeping. Therefore, the overhead is reduced by $(1-R)\times100\%$.
To perform this three-fold joint optimization, we also exploit the bandit-based MCTS approach. We first decide the beam sweep period. Then we decide the beamwidth level, which gives the number of sectors that cover the whole space, denoted by $N_\text{s}$. Finally we select $N_\text{s}R$ beams of $N_\text{s}$. The decision-making processes on the first two sub-actions are the same as presented in Fig.~\ref{fig:MTCS}. The only difference is the extra step that decides which $N_\text{s}R$ beams are to be tested. 
Unlike the traditional MAB framework where a single arm is picked each time, our scenario requires choosing $N_\text{s}R$ arms, which is known as a best-$K$ identification problem in the bandit literature. 
The beam selection and reward backpropagation in MTCS is shown in Fig.~\ref{fig:topKbeam} and further explained as below:
Each beam has its own UCB index (or posterior distribution if TS is used). Among all candidate beams, the $N_\text{s}R$ beams associated with the $N_\text{s}R$ largest UCB (or predicted mean if TS is used) are used for beam sweeping. 
After the beam sweeping process, the link quality of each beam is sent back to the transmitter. These feedbacks of beams would be used to update their UCB index (or posterior distribution if TS is used). The beam that provides the highest SNR is used for data transmission and the average effective data rate throughout the time slot is backpropagated to their ancestor nodes.

The evaluation results are given in Fig~\ref{fig:tslot_bw_beam_selection}. As we can see, system throughput can be significantly improved in our studied case. For example, by selecting 1/2 of the candidate beams ($R=1/2$), the bandit-based MCTS policy can improve the system data rate by more than 20\% in comparison with the case that $R=1$. The improvement over a random selection policy that enumerates all candidate beams is even more than 150\%.
Although the performance improvement is promising, the parameter-tuning over $R$, namely deciding how much overhead to be reduced is crucial and tricky. One potential future direction would be to further expand this three-fold joint optimization to a four-fold one. Besides, it might be interesting to learn a set of beams that are not necessarily of the same width. This is challenging as the action space would surge up. It could be helpful to explore the correlation among the performance of beams that have different beamwidth but point to the same directions.

In this section, we have showcased three concatenated scenarios of applying MAB in solving the mmWave link configuration problems, which reveals that:
(1) MAB is feasible for its fast convergence and computational efficiency. The performance improvement in terms of system data rate is significant.
(2) Bandit-based MCTS can factor a large action space into several sequential ones such that the optimal configuration could be learned in a fast way.

\section{Future research directions}
There are many promising research directions about applying MAB in mmWave systems. We highlight some of them below.

\subsection{Time-varying MAB for non-stationary system}
In this article, we have focused on applying the stationary MAB framework into mmWave systems, where we treat that the reward distributions are stationary for simplicity. The underlying randomness in wireless systems could be time-varying. Accordingly, it is important to further design bandit algorithms that can specifically handle the non-stationary reward. For example, a parallel subroutine algorithm could be designed to detect whether the underlying system has a significant change, and then the classical stationary MAB algorithms would be reset when such a scenario is detected. Besides, if the reward distributions evolve smoothly over time, a sliding-window approach that only uses the most recent observations for decision-making might be helpful.
It is imperative to design a unified bandit algorithm that automatically handles both abruptly and smoothly changing environments.

\subsection{Double-side link configuration}
The MAB framework and its applications in mmWave systems presented in this article only consist of a single agent (aka decision-maker). 
In wireless systems, lots of link configuration parameters exist in both transceivers. For example, both the BS and users could configure their own beamwidth for beam sweeping. One naive solution to this problem is to extend the number of arms to $K_\text{T} \times K_\text{R}$, where $K_\text{T}$ ($K_\text{R}$) is the number of arms the transmitter (receiver). This approach however increases the algorithm complexity quadratically. Is
there any possibility to set up two agents learning their own MAB models and might exchange information to perform co-decision making? This is an open question.

\subsection{Federated MAB for high communication efficiency and data privacy}
The examples given in this article only showcased the application of MAB in the downlink transmission scenarios. Therefore, it is interesting to extend them for multi-user uplink transmission, where each user has its own local MAB model running for the link configuration. Meanwhile, they can share with each other their learned model through the BS or device-2-device connection. This idea leads to the design of federated MAB algorithm~\cite{federate_MAB_Shen_21}, which is inspired by the principle of federated learning: training a model across multiple decentralized agents and merging the locally trained models in a communication-efficient and data-private way. Another related challenge to federated MAB is the adversarial users who could attack the systems by sharing malicious local models, which calls for designing defense mechanisms to protect the global model. 

\section{Conclusions}\label{sec:discussion}
Efficient operation of mmWave wireless networks requires careful selection of system parameters from among many possibilities. Reinforcement learning is one approach that can be used to adaptively select those parameters in a way that is data-driven. In this article, we explained the key idea of MABs and how they can be used to solve certain mmWave link configuration problems related to beam training. We found that such algorithms were able to dynamically learn the optimal mmWave link configuration strategies in a site-specific or user-specific manner. The MAB approach has advantages over other strategies like deep reinforcement learning in that it has fast convergence and low computational complexity make it more suitable for the small timescales of mmWave communication. We outlined several directions for future research that will push MABs to be better customized for the specific problems faced by next generation wireless problems. 

\bibliographystyle{IEEEtran}
\bibliography{ref-mag.bib}

\begin{IEEEbiographynophoto}
	{Yi Zhang}
	(S'15-M'21) received his Ph.D. degree in Electrical and Computer Engineering from The University of Texas at Austin in 2021. Before that, he received B.S./M.S. degrees from Xi'an Jiaotong University in 2014 and 2017, respectively. He also received an Engineer's degree from Ecole Centrale de Nantes, France, in 2017. His research interests include wireless, networking, signal processing, reinforcement learning (DRL and bandit), mmWave, deep learning-based wireless communication, and wireless system prototyping.
\end{IEEEbiographynophoto}

\begin{IEEEbiographynophoto}
	{Robert W. Heath Jr.}
	(S'96-M'01-SM'06-F'11)  received the B.S. and M.S. degrees from the University of Virginia, Charlottesville, VA, in 1996 and 1997 respectively, and the Ph.D. from Stanford University, Stanford, CA, in 2002, all in electrical engineering. 
	He is the Lampe Distinguished Professor at North Carolina State University. From 2002-2020 he was with The University of Texas at Austin, most recently as Cockrell Family Regents Chair in Engineering and Director of UT SAVES. He is also President and CEO of MIMO Wireless Inc. He authored ``Introduction to Wireless Digital Communication'' (Prentice Hall, 2017) and ``Digital Wireless Communication: Physical Layer Exploration Lab Using the NI USRP'' (National Technology and Science Press, 2012), and co-authored ``Millimeter Wave Wireless Communications'' (Prentice Hall, 2014) and ``Foundations of MIMO Communication'' (Cambridge University Press, 2018). He is currently Editor-in-Chief of IEEE Signal Processing Magazine. 
\end{IEEEbiographynophoto}

\end{document}